\documentclass[iop]{emulateapj}
\usepackage{apjfonts}

\usepackage{natbib}
\usepackage{color}

\slugcomment{Draft \today}

\shorttitle{Double Pulsar Eclipses I}
\shortauthors{Breton et al.}

\begin{document}

\title{The Double Pulsar Eclipses I: Phenomenology and Multi-frequency Analysis}

\author{R. P. Breton\altaffilmark{1,2}, V. M. Kaspi\altaffilmark{2}, M. A. McLaughlin\altaffilmark{3,4}, M. Lyutikov\altaffilmark{5}, M. Kramer\altaffilmark{6}, I. H. Stairs\altaffilmark{7}, S. M. Ransom\altaffilmark{8}, R. D. Ferdman\altaffilmark{9}, F. Camilo\altaffilmark{10}, {\sc and} A. Possenti\altaffilmark{11}}

\altaffiltext{1}{Department of Astronomy and Astrophysics, University of Toronto, Toronto, ON M5S 3H4, Canada; breton@astro.utoronto.ca}
\altaffiltext{2}{Department of Physics, McGill University, Montreal, QC H3A 2T8, Canada}
\altaffiltext{3}{Department of Physics, West Virginia University, Morgantown, WV 26506, USA}
\altaffiltext{4}{National Radio Astronomy Observatory, Green Bank, WV 24944, USA}
\altaffiltext{5}{Department of Physics, Purdue University, West Lafayette, IN 47907, USA}
\altaffiltext{6}{Max Planck Institut f\"ur Radioastronomie, Auf dem H\"ugel 69, 53121 Bonn, Germany}
\altaffiltext{7}{Department of Physics and Astronomy, University of British Columbia, Vancouver, BC V6T 1Z1, Canada}
\altaffiltext{8}{National Radio Astronomy Observatory, Charlottesville, VA 22903, USA}
\altaffiltext{9}{University of Manchester, School of Physics and Astronomy, Jodrell Bank Centre for Astrophysics, Alan Turing Building, Manchester, M13 9PL, UK}
\altaffiltext{10}{Columbia Astrophysics Laboratory, Columbia University, New York, NY 10027, USA}
\altaffiltext{11}{INAF Osservatorio Astronomico di Cagliari, Poggio dei Pini, strada 54, I-09012 Capoterra, Italy}

\begin{abstract}
The double pulsar PSR~J0737$-$3039A/B displays short, 30 s eclipses that arise around conjunction when the radio waves emitted by pulsar A are absorbed as they propagate through the magnetosphere of its companion pulsar B. These eclipses offer a unique opportunity to probe directly the magnetospheric structure and the plasma properties of pulsar B. We have performed a comprehensive analysis of the eclipse phenomenology using multi-frequency radio observations obtained with the Green Bank Telescope. We have characterized the periodic flux modulations previously discovered at 820\,MHz by \citeauthor{mcl04a}, and investigated the radio frequency dependence of the duration and depth of the eclipses. Based on their weak radio frequency evolution, we conclude that the plasma in pulsar B's magnetosphere requires a large multiplicity factor ($\sim 10^5$). We also found that, as expected, flux modulations are present at all radio frequencies in which eclipses can be detected. Their complex behavior is consistent with the confinement of the absorbing plasma in the dipolar magnetic field of pulsar B as suggested by \citeauthor{lyu05a} and such a geometric connection explains that the observed periodicity is harmonically related to pulsar B's spin frequency. We observe that the eclipses require a sharp transition region beyond which the plasma density drops off abruptly. Such a region defines a plasmasphere which would be well inside the magnetospheric boundary of an undisturbed pulsar. It is also two times smaller than the expected standoff radius calculated using the balance of the wind pressure from pulsar A and the nominally estimated magnetic pressure of pulsar B.
\end{abstract}


\section{Introduction}\label{s:intro}

The double pulsar PSR~J0737$-$3039A/B is one of the most celebrated systems in astrophysics. On the one hand, the fact that it comprises two neutron stars, both visible as radio pulsars, combined with being the most relativistic pulsar binary, allows for the measurement of post-Keplerian parameters at an unprecedented level of accuracy \citep{kra06a}. Its short coalescence timescale significantly increases the estimates of the expected double neutron star merging rate in galaxies like ours \citep{bur03b}, the timing provides the most precise test of general relativity in the strong-field regime so far \citep{kra06a} and, more recently, a high-precision measurement of the relativistic spin precession of pulsar B yielded the first theory-independent constraint on spin-orbit coupling in the strong-field regime \citep{bre08a}.

On the other hand, the double pulsar offers a unique laboratory to study a variety of pulsar phenomena. This is related to the fact that pulsar ``A'', the 23 ms spin period recycled pulsar, has an energetic wind roughly 3625 times stronger than that of its 2.8 s companion, ``B''. As a consequence, the dynamic pressure from pulsar A's wind confines and distorts the magnetosphere of pulsar B, whose radio pulsations are difficult to detect throughout most of the orbit except in two bright regions \citep{lyn04a}. It was proposed that pulsar A's wind changes the direction of B's radio emission in different parts of the orbit \citep{lyu05b}, but the exact nature of the phenomenon still remains uncertain despite phenomenological studies \citep{bur05a,per10b}. Pulsar B also shows sub-pulse drifting at the spin frequency of pulsar A in one of the bright orbital phase windows \citep{mcl04b}, demonstrating a direct modulation of B's emission by the non-axisymmetric dynamic pressure of A's Poynting-dominated wind \citep{spit06a}. It might even be possible to use this effect as an extra tool to time the system \citep{fre09a}.

Arguably, the most impressive display of the interaction between the two pulsars are the radio eclipses of pulsar A when it passes behind pulsar B, thanks to the fortunate nearly edge-on viewing geometry. As opposed to eclipses in other types of pulsar binaries, such as those resulting from the stellar wind of a main sequence companion \citep[e.g.][]{joh94a} or the ablated material from the surface of a low-mass companion \citep[e.g.][]{tho91a,sta01a}, those of the double pulsar have a magnetospheric origin. The duration of the eclipses implies a sliced transverse size of 18,100\,km if the absorbing region lies around pulsar B at a distance of 2.9\,lt-s from pulsar A \citep{kas04a}; this is well within pulsar B's inferred light cylinder radius, which is 135,000\,km, as well as the estimated 40,000\,km magnetosphere radius in the presence of pulsar A's wind. This first occurrence of eclipses in a double neutron star system provides an exceptional way of probing the magnetosphere of a pulsar.

One of the most outstanding problems in understanding pulsars is the generation of coherent radio emission across a wide range of rotational periods and magnetic fields. The physics behind the emission is intimately linked to the properties of their magnetospheres, in particular their structure and the particle density, which have yet to be well constrained observationally and theoretically. A realistic magnetosphere may deviate from a classical \citet{gj69} dipolar magnetosphere in having a particle density exceeding the so-called Goldreich$-$Julian density, except very close to its surface \citep{rude75a}. Theoretical work shows that the creation of additional pair plasma in the magnetosphere is a key ingredient to power the radio emission but progress still has to be made to reconcile theory and observations \citep{aron00a}. In the case of the double pulsar, we have for the first time a powerful source of radio light positioned behind, albeit externally modified, a pulsar magnetosphere.

As pointed out by \citet{raf05a}, resonant cyclotron absorption should be the dominant source of extinction of radio waves in an isolated pulsar magnetosphere but this mechanism would only yield optical depths in the range of unity if the multiplicity factor is the order of 100. One way to reach such a value is if the particle number density and trapping efficiency can be increased via the interaction with pulsar A. As a consequence, though, particles will be heated to relativistic velocities and resonant synchrotron absorption will inevitably become the main source of opacity. While resonant synchrotron absorption appears to be a viable eclipse mechanism, a number of questions still remain. How large can the electron multiplicity factor become? What is the radial density profile of the electrons? Through which mechanism is the heating resulting from the interaction with pulsar A operating?

In their model, \citet[hereafter RG05]{raf05a} propose that if the illumination from pulsar A's radio beam can pump sufficient perpendicular momentum in the particles of pulsar B's magnetosphere via cyclotron absorption that the trapping efficiency can increase significantly. In turn, the particle density would be able to grow and reach a steady state. Their model predicts a multiplicity factor $\sim 100$ at a radius 10,000\,km for fiducial neutron star parameters. It also predicts an eclipse duration scaling as $\nu^{-1/3}$, where $\nu$ is the observed radio frequency.

In contrast, the eclipse model from \citet[hereafter LT05I]{lyu05a}, suggests that the multiplicity factor is much larger, $\sim 10^5$. They believe it can be provided by torsional Alfv\'en waves resulting from the interaction of pulsar B's magnetosphere with pulsar A's relativistic wind. A particular feature of their model is that particles should be trapped in the closed field lines of the magnetosphere and have a roughly constant density. To simplify the model further, they suggest that the density distribution can be assumed constant up to the field lines closing at some characteristic radius, beyond which trapping becomes inefficient and the density drops to negligible values. Alternatively, they also suggest a different version of their model, hereafter LT05II, that retains the overall geometrical properties but in which the absorbing electrons are supported by an ion cloud. In this model, the electron number density is constant along each field line and proportional to the electron temperature, which is itself a function of the radius (suggested to be a power law, $T_e(r) \propto r^7$).

In this paper, we revisit some of the phenomenological properties of the double pulsar eclipses in light of the progress made in our understanding of the phenomenon using the LT05I model. Indeed, this model, which relies on a magnetic dipole misaligned with respect to the spin axis, has been used successfully to reproduce the eclipse light curves by \citet{bre08a} and to explain the B's pulse profile changes by \citet{per10b}. After introducing our observations and data reduction in Section~\ref{s:observations}, we present in Section~\ref{s:phenomenology} a multi-radio-frequency analysis of the peculiar eclipse properties including flux modulations, their periodicity, duration and depth. Whereas the periodic behavior of the flux modulations is tightly connected to the geometry of pulsar B with respect to the orbit, we show that the eclipse duration and depth can be used to probe the properties of the plasma confined in the magnetosphere. In Section~\ref{s:conclusion}, we summarize our work and provide some concluding remarks.

\section{Observations and Data Reduction}\label{s:observations}

We observed the double pulsar regularly since 2003 December as part of a multi-purpose monitoring campaign aimed primarily at the timing of the pulsars, but also at the investigation of the different phenomena displayed by this system such as the eclipses of pulsar A, the orbital modulation of pulsar B's flux, and its subpulse drifting. Data used for the analysis presented in this paper were acquired at the Green Bank Telescope (GBT), in West Virginia, with the SPIGOT and BCPM backend instruments \citep{kap05a, bac97a} between 2003 December and 2007 November (proposal codes GBT03C$-$041, GBT04B$-$026, GBT05B$-$042, GBT06B$-$018 and GBT07B$-$029). For the multi-frequency analysis reported in the next section, we use only the observations obtained during the first year (MJDs 52984$-$53379) because they cover a large range of radio frequencies while also spanning a short enough time that the eclipse profile did not change significantly due to the effect of relativistic precession of pulsar B's spin axis. Relevant details about the observations and setup are presented in Tables \ref{t:data} and \ref{t:setup}.

\begin{table}
\caption{Summary of GBT Observations}\label{t:data}
\footnotesize
\begin{center}
\begin{tabular}{cccc} \hline
Date {\scriptsize (MJD)} & Frequency {\scriptsize (MHz)} & Backend & Number of Eclipses \\ \hline
52984 & 1400 & BCPM & 2 \\
52992 & 820 & BCPM & 2 \\
52996 & 2200 & BCPM & 2 \\
52997 & 820 & SPIGOT & 3 \\
53005 & 427 & BCPM & 1 \\
53005 & 427 & SPIGOT & 2 \\
53191 & 325 & SPIGOT & 2 \\
53211 & 820 & SPIGOT & 2 \\
53311 & 820 & SPIGOT & 3 \\
53212 & 1950 & SPIGOT & 2 \\
53378 & 1950 & SPIGOT & 2 \\
53379 & 820 & SPIGOT & 3 \\
54199 & 820 & SPIGOT & 1 \\
54201 & 820 & SPIGOT & 2 \\
54202 & 820 & SPIGOT & 2 \\
54204 & 820 & SPIGOT & 3 \\
\hline
\end{tabular}
\end{center}
\end{table}

\begin{table}
\caption{GBT Observing Setup}\label{t:setup}
\footnotesize
\begin{center}
\begin{tabular}{cccc} \hline
Frequency & Bandwidth & Number of & Sampling \\
{\scriptsize (MHz)} & {\scriptsize (MHz)} & Channels & {\scriptsize ($\mu$s)} \\ \hline
\multicolumn{4}{c}{BCPM} \\ \hline
427 & 48 & 96 & 72 \\
820 & 48 & 96 & 72 \\
1400 & 96 & 96 & 72 \\
2200 & 96 & 96 & 72 \\
\hline
\multicolumn{4}{c}{SPIGOT} \\ \hline
325 & 50 & 1024 & 81.92 \\
427 & 50 & 1024 & 81.92 \\
820 & 50 & 1024 & 81.92 \\
1950 & 600 & 1024 & 81.92 \\
\hline
\end{tabular}
\end{center}
\end{table}

We performed the initial data reduction using the pulsar analysis packages {\tt PRESTO}\footnote{{\tt PRESTO} is freely available at \url{http://www.cv.nrao.edu/~sransom/presto}.} \citep{ran02a} and {\tt SIGPROC}.\footnote{D. R. Lorimer. {\tt SIGPROC} is freely available at \url{http://sigproc.sourceforge.net}.} First, we dedispersed the data (DM = 48.920 cm$^{-3}$\,pc) to correct for the frequency-dependent travel time in the ionized interstellar medium. Then we generated folded data products using the predicted spin periods of pulsars A and B. For these two steps, we used the timing solution presented in \citet{kra06a}.

We generated a pulsed flux time series for each segment of observations containing an eclipse of pulsar A in the same way as presented in \citet{bre08a}. We generated high signal-to-noise (S/N) pulse templates of pulsar A by integrating its flux over each observation. Then we calculated the relative pulsed flux density of pulsar A by fitting the corresponding pulse profile templates to folded intervals consisting of the sum of four individual pulses of pulsar A ($\sim$91\,ms). This choice provides a good balance between S/N and time resolution of the eclipse features. Note that as opposed to {\tt SIGPROC}, which folds at an integer number of pulses, {\tt PRESTO}'s folding algorithm is based on the data size and hence despite the fact that the exact number of pulses is generally four, there are a handful of data points consisting of three pulses only. We renormalized these data points to correct for their shorter integration time. Finally, we normalized the time series such that the flux average in the out-of-eclipse region is unity.

Radio frequency interference is an issue that limits the quality of our data. We observe that, with similar setup and observing conditions, the S/N can vary substantially from one observation to another depending on the level of interference. Some of it can be mitigated if it is well localized in the radio spectrum but our determination of the pulsed flux amplitude is mainly affected by short interference ``bursts'', which are impossible to eliminate, especially because we deal with short folding intervals. Generally speaking, GBT observations at 820\,MHz are relatively free of interference and this motivates why data are often recorded at this frequency.

Using the timing solution of \citet{kra06a}, we calculated the orbital phase at each data point of our flux time series. Throughout this paper, we refer to the orbital phases using the superior conjunction of pulsar A (i.e. the eclipse mid-point) as zero-point. We also empirically determined the spin phases of pulsar B associated with the data points of pulsar A's time series using the method described in Section~1 of the Supporting Online Material of \citet{bre08a}.

\section{Eclipse Phenomenology}\label{s:phenomenology}
\subsection{Modulations}
\begin{figure}
\centering
\includegraphics[width=0.95\columnwidth]{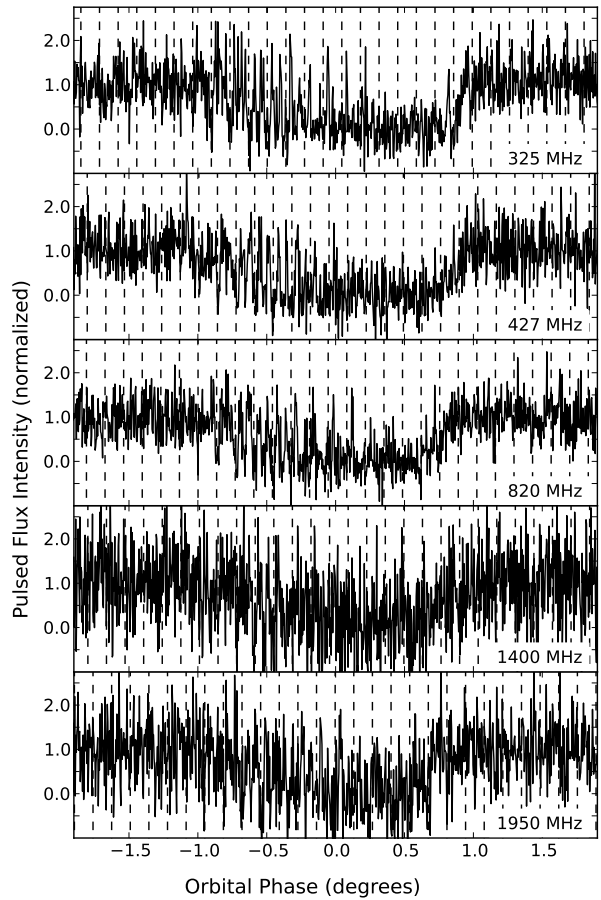}
\caption{Typical individual eclipse light curves of pulsar A observed with the GBT at 325, 427, 820, 1400, and 1950\,MHz. Each point in the light curves was made of four individual pulses from pulsar A, while the vertical dashed lines mark each one of pulsar B's pulsations. At the three lowest radio frequencies, the modulation behavior is clearly visible whereas the poorer S/N at higher frequencies makes it hard to identify unambiguously. Note that these observations present higher-than-normal S/N, mainly due to the level of radio frequency interference being low.}
\label{f:eclipse}
\end{figure}

What makes the double pulsar eclipses very different from any other eclipses observed to date are the rapid flux variations discovered by \citet{mcl04a} in a high time resolution analysis of 820\,MHz eclipse light curves. They found that not only does pulsar A's flux vary on short timescales, but that it is also modulated in phase with the rotation of pulsar B (see Figure \ref{f:eclipse}). This kind of behavior has considerable importance for constraining the eclipse origin. Early theoretical work to explain the eclipses involved the magnetosheath of pulsar B, which would form in the direction opposed to pulsar A as a result of the relativistic shock of its wind with pulsar B's magnetosphere \citep{aro05a,lyu04a}. This family of models, however, can now be excluded as they are unable to reproduce the rapid modulation seen in the eclipses.

\subsubsection{Periodicity}\label{s:periodicity}
Although the connection between the flux variability and the rotational phase of pulsar B has been established at 820\,MHz \citep{mcl04a}, the presence of such behavior at other radio frequencies, although expected, is yet to be confirmed. For intrinsic (i.e. spectrum) and technical (i.e. instrument bandwidth, sensitivity and radio interference) reasons, pulsar A at GBT is generally more strongly detected at 820\,MHz than in other frequency bands. Hence, most of the data were collected at this frequency, which was also chosen for the original analysis by \citet{mcl04a}.

\begin{figure}
\centering
\includegraphics[width=0.95\columnwidth]{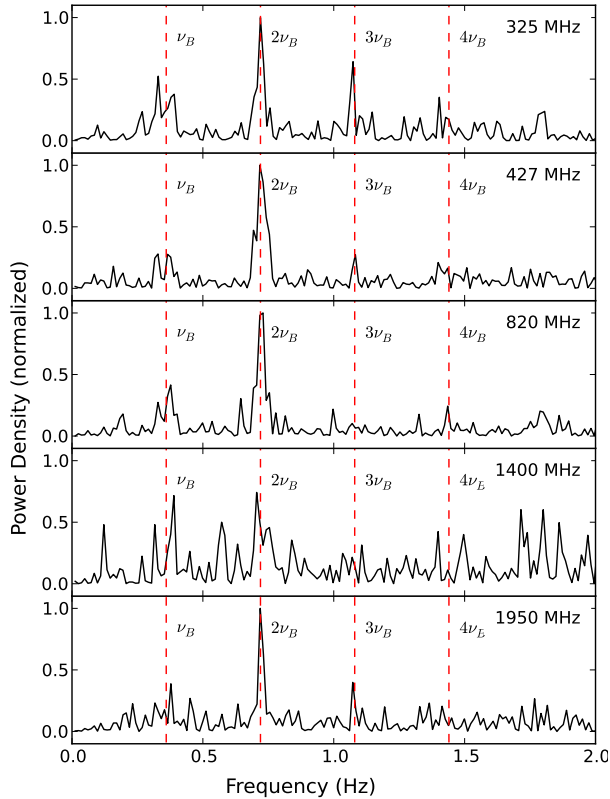}
\caption{Power spectra of the eclipse light curves presented in Figure \ref{f:eclipse}. Data were high-pass filtered in order to remove the low-frequency eclipse trend. The vertical dashed lines indicate the first four harmonics of pulsar B's spin frequency ($\nu_{\rm B}$) at 0.36, 0.72, 1.08, and 1.44\,Hz, respectively.}
\label{f:powerspectra}
\end{figure}

We performed a Fourier analysis on pulsar A's flux lightcurves in order to improve the sensitivity to detect modulations at other frequencies, as well as to obtain quantitative information about this behavior. First, we removed the overall low-frequency trend of the eclipse using a high-pass filter consisting of two running means having a window size of 71 data points (i.e., 0.3$^\circ$ in orbital phase). We found that the power spectra of the individual light curves observed at radio frequencies 325, 427, 820, 1400, and 1950\,MHz show significant excess power at the frequency and higher harmonics of the rotational period of pulsar B (see Figure \ref{f:powerspectra}). These features are easily visible even in data presenting poor S/N and for which the modulation is not clearly identifiable by eye.

The periodic nature of the modulations, harmonically related to the spin frequency of pulsar B, is important because it suggests a strong connection between the distribution of the absorbing material and the geometry of pulsar B. We also searched for modulations at radio frequencies higher than 1950\,MHz but the very poor S/N --- at 2200\,MHz, S/N$\sim 0.25$ per data point in the out-of-eclipse region --- prevented us from seeing either an eclipse trend or finding any significant feature in the power spectrum. Unfortunately, pulsar A becomes so faint at higher radio frequencies that in order to obtain a reasonable S/N, the binning of the sub-fold intervals must be longer than the modulation timescale.

\begin{figure}
\centering
\includegraphics[width=0.95\columnwidth]{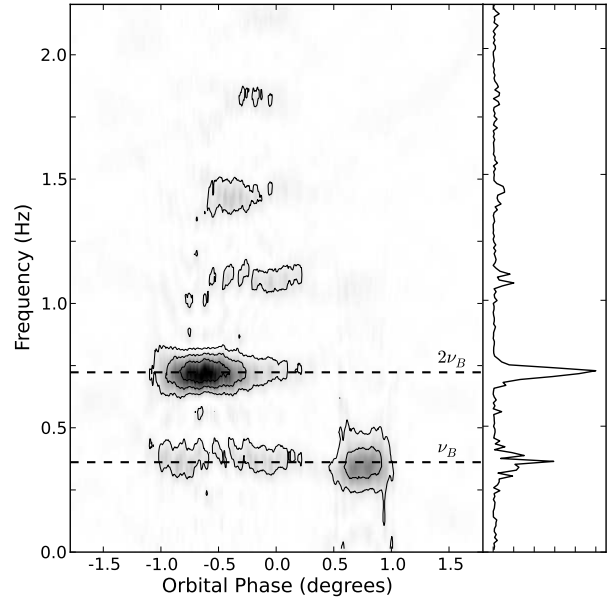}
\caption{Dynamic power spectrum analysis of the average light curve consisting of eight eclipses observed at epoch MJD 54200. In this figure, the frequency of the dynamic power spectrum is displayed vertically as a function of orbital phase, horizontally. The color intensity indicates the spectral energy density, with black being the highest value, and three isocontour levels drawn with solid lines. The right panel shows the ``standard'' power spectrum averaged over orbital phases $-2^\circ$ to $2^\circ$. Data were high-pass filtered in order to remove the low-frequency eclipse trend. The horizontal dashed lines indicate the first two harmonics of pulsar B's spin frequency ($\nu_{\rm B}$) at 0.36 and 0.72\,Hz, respectively.}
\label{f:dynamic}
\end{figure}

Another insightful way to analyze the flux modulations is through the use of a dynamic (also known as windowed) Fourier transform. This technique consists of calculating the Fourier transform within subsections of a time series and repeating the process after translating the window kernel. The corresponding power spectra can then be calculated and stacked in a two-dimensional plot so that the dynamic information about the time evolution of the power content can be visualized. Figure \ref{f:dynamic} shows such a dynamic power spectrum obtained from the average of eight 820\,MHz observations observed around MJD 54200. We chose these data because they comprise the largest sequence of 820\,MHz data that we obtained over a short time span. The averaging of the light curves was done as described in \citet{mcl04a}. That is, because the phase of the modulation varies from one eclipse to another, the light curves were shifted by up to $\pm P_{\rm spin,B}$ in order to sum the features coherently.

Our dynamic power spectrum analysis yields two important results. First, no periodic flux variations are detected out of the eclipse region, where the power content is well characterized by white noise. Despite the fact that our pulsed flux time series are made from data folded at pulsar A's spin period, a process which should average out the signature of pulsar B, the fact that modulations occur only during the eclipse confirms they are not an artifact of the data reduction. Second, we find that the power content significantly evolves during the eclipse. As seen in Figure \ref{f:dynamic}, we observe that the eclipse ingress is initially characterized by modulations having significant power content at the spin frequency ($\nu_{\rm B}$) and the second harmonic ($2\nu_{\rm B}$) of pulsar B. Then, the modulation switches to a mode dominated by the second harmonic ($2\nu_{\rm B}$) around orbital phase $-0.5^\circ$. It then moves to a mode characterized by a high harmonic content, which indicates that the modulation features are very narrow. Finally, the modulations temporarily stop before resuming in the first harmonic ($\nu_{\rm B}$) at the egress.

\subsubsection{Folded Light Curve}
\begin{figure}
\centering
\includegraphics[width=0.95\columnwidth]{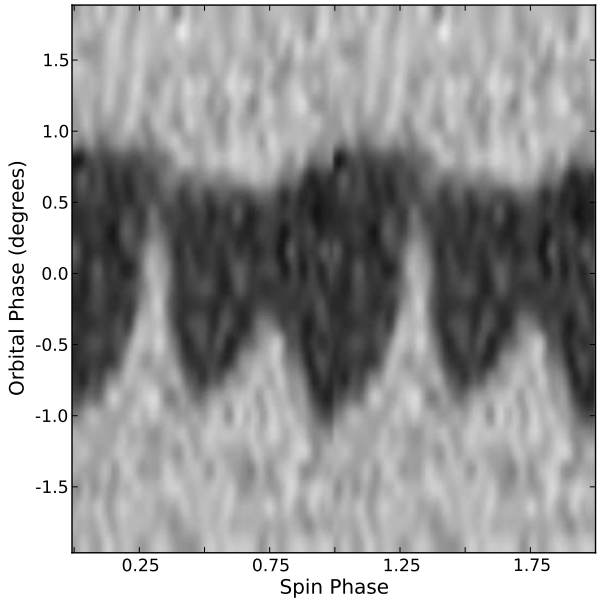}
\caption{Folded eclipse profile for the eight eclipses observed at epoch MJD 54200. Eclipses were combined before being folded at the spin period of pulsar B. Lighter regions have higher flux intensity levels.}
\label{f:fold}
\end{figure}

To complete the picture, folding the light curve at the spin period of pulsar B provides one more way to visualize the properties of the eclipses, in particular the dynamic information of the phase of the modulations. Figure \ref{f:fold} shows the same eclipse data as in Figure \ref{f:dynamic} but without the high-pass filtering. As pointed out by \citet{mcl04a}, we can clearly see that the eclipse is longest when the spin phase of pulsar B is 0.0 (i.e., when we receive its radio pulse). In a narrow range around pulsar B's spin phase 0.25, pulsar A barely disappears. We also observe that initially, most of the modulation power is due to the absorption feature located around spin phase 0.0, which then also appears around spin phase 0.5. Toward the center of the eclipse, at orbital phase 0.0, the modulation consists mainly of narrow emission features at B pulsar phase 0.25.

\begin{figure*}
\centering
\includegraphics[width=0.4\hsize]{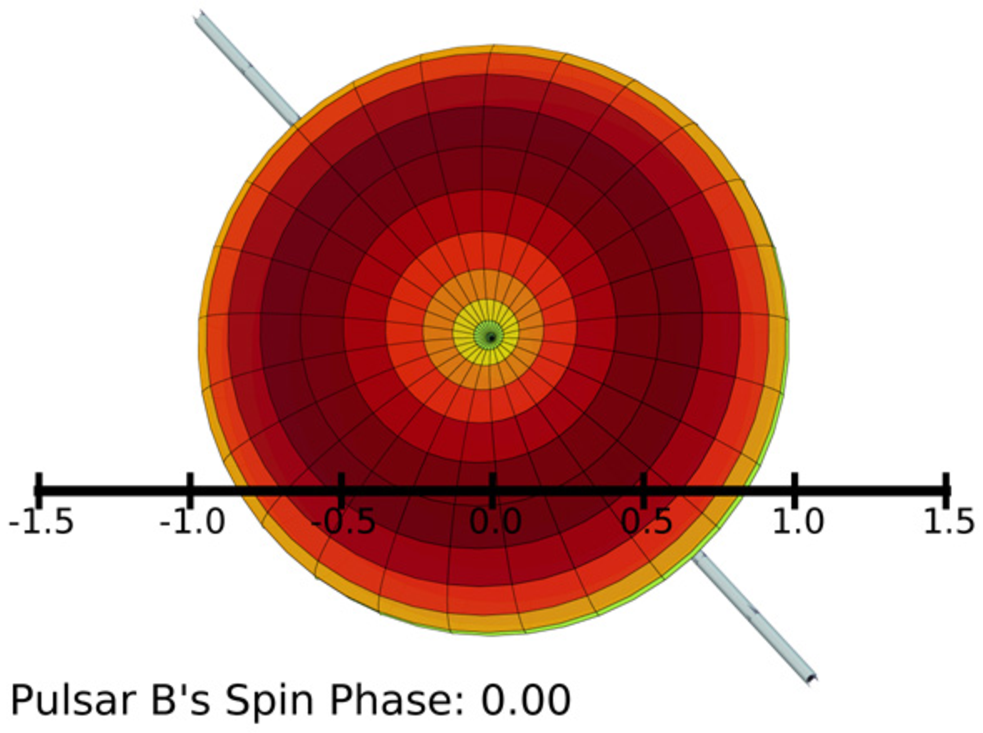}
\includegraphics[width=0.4\hsize]{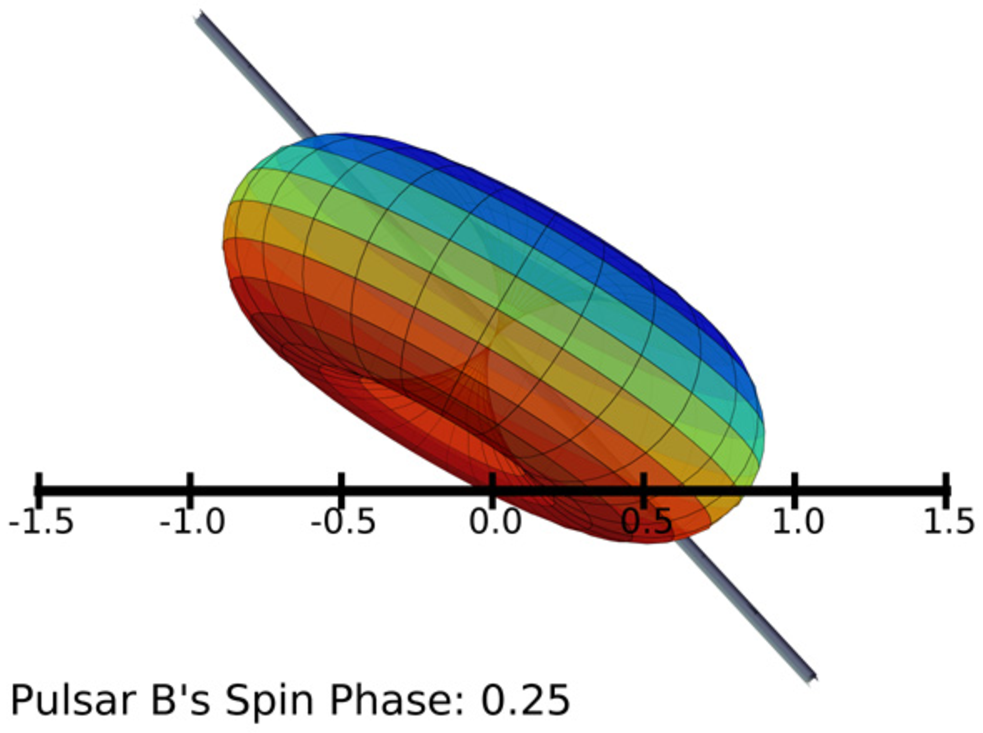}
\includegraphics[width=0.4\hsize]{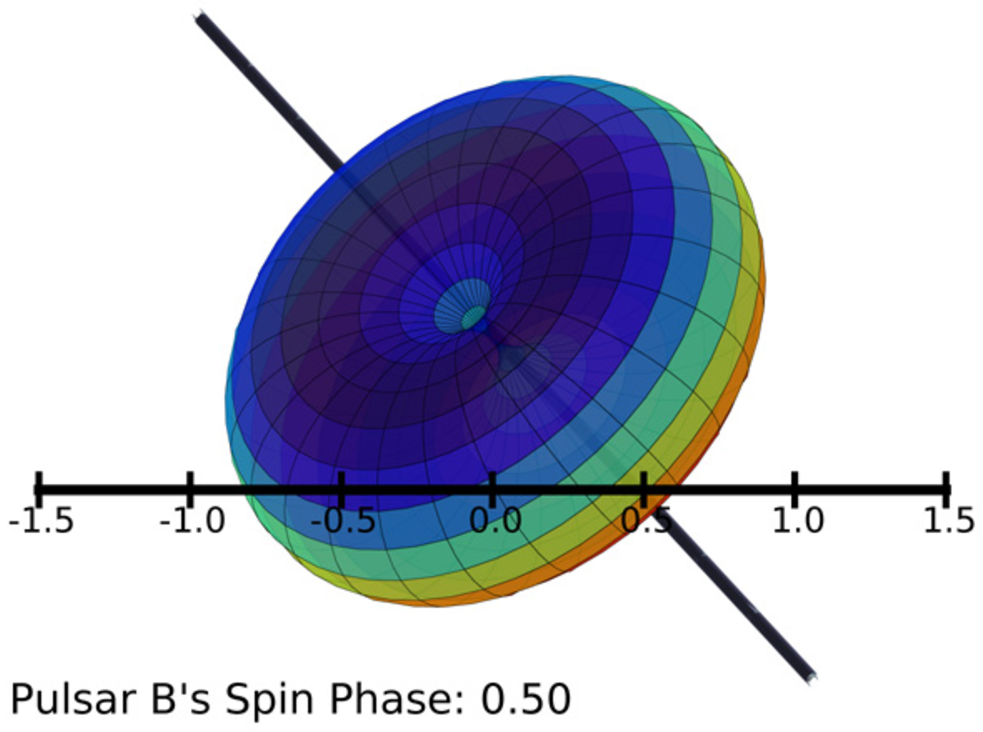}
\includegraphics[width=0.4\hsize]{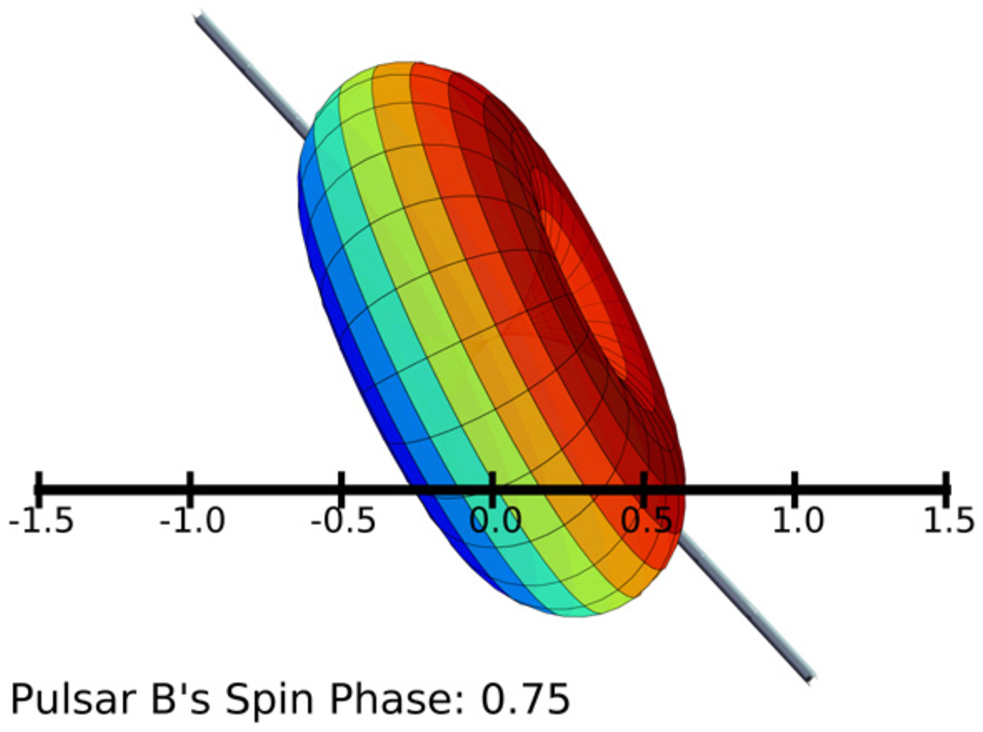}
\caption{Illustration of the truncated magnetosphere of pulsar B within which the absorbing plasma is confined as seen from an observer located on Earth. The horizontal line represents the projected motion of pulsar A, which goes from left to right, with the labels indicating the orbital phases in degrees. The system configuration is that of the best-fit geometry from \citet{bre08a}, with the spin axis, depicted as a gray rod, tilted $130^\circ$ with respect to the orbital plane (vertical, in the plane of the page), and oriented $51^\circ$ away from our line-of-sight vector (pointing directly off the page). The angle between the spin and magnetic axes is $71^\circ$. The four panels display different spin phases of pulsar B, with the upper left corresponding to pulsar B's spin phase 0.0 (i.e., when pulsar B's radio emission is illuminating us). Depending on the location of pulsar A, the rotation of the magnetosphere of pulsar B can modulate the flux from A twice per rotation at the ingress (leftmost side), or only once closer to the conjunction time (center). }
\label{f:donut}
\end{figure*}

The complex modulation pattern and the asymmetric shape of the eclipse arise naturally as a consequence of the magnetic dipole being misaligned with respect to the spin axis in the context of the LT05I geometrical model. As seen in Figure \ref{f:donut}, if we consider the closed field lines reaching maximum extent, the magnetosphere of pulsar B resembles a torus which presents its largest cross-section at spin phase 0.0, when its radio beam is illuminating us. Conversely, the narrowest side of the dipole is visible about a quarter of a rotation later/before. At the start of the eclipse, the magnetosphere of pulsar B intercepts pulsar A's line of sight twice per rotation while it remains visible to us when the magnetosphere is viewed from the side. Gradually, as the orbital motion brings pulsar A closer along the line of sight of the projected magnetosphere of pulsar B, A becomes hidden most of the time and barely emerges in a narrow transparent window when the magnetosphere is viewed sideways. The offset of the magnetic axis with respect to the spin axis, $\sim 20^\circ$ \citep{bre08a}, is such that the magnetosphere of pulsar B will only clear out the line of sight at spin phase 0.25, but not at spin phase 0.75, which explains the single modulation per rotation of B in the light curve around conjunction. The large ingress/egress asymmetry appears from the fact that the longitude of pulsar B's spin axis, about $50^\circ$ away from our line-of-sight vector in 2004, causes the magnetosphere to intercept the line of sight only during the egress for the spin phases 0.25 and 0.75. Based on the geometry inferred from the truncated dipole LT05I model, relativistic spin precession of pulsar B should bring its spin axis toward our line of sight and the two should lie in a plane perpendicular to the orbit around year 2013. At this time, the eclipse profile should be nearly symmetrical.

\subsection{Duration}
In principle, the eclipse profile not only offers a way of determining the geometry of pulsar B but also of studying the properties of the plasma inside its magnetosphere. For instance, the radial distribution of the absorbing particles should change the amplitude and shape of the eclipse features. However, the fact that the optical depth increases very abruptly to large values at the eclipse characteristic radius makes it hard to probe the inner regions of the magnetosphere since most of the differences that arise in the light curve profile happen on a timescale and at a flux level that is beyond the observational capabilities of the GBT.

On the other hand, the behavior of the eclipse light curves as a function of radio frequency might provide an easier alternative to learn about the plasma. \citet{kas04a} reported that the eclipse duration is only slightly frequency dependent in the range $427-1400$\,MHz. From the full width at half-maximum (FWHM) of the eclipses, which they calculated after fitting a Fermi function to each half of the eclipse, they found a small, non-zero linear decrease of duration as a function of frequency ($-4.52(3) \times 10^{-7}$ orbits\,MHz$^{-1}$, i.e., $-1.63(1) \times 10^{-4}\,^\circ$\,MHz$^{-1}$). Since our data set includes observations made over a larger range of radio frequencies, $350-1950$\,MHz,\footnote{Data at higher frequencies could not be used because they do not have sufficient S/N to detect the eclipses.} we conducted a similar analysis in order to test the predictions of different eclipse models.

We chose observations made around the same epoch, between 2003 December and 2005 January (MJDs $52984-53378$), in order to reduce effects of the long-term evolution of the eclipse profile due to the relativistic spin precession of pulsar B \citep{bre08a}. We fitted the high time resolution lightcurves to a double Fermi function:
\begin{equation}\label{eq:fermi}
    F(\phi) = \beta \left[ \frac{1}{e^{(\phi+\phi_i)/w_i}+1} + \frac{1}{e^{(\phi+\phi_e)/w_e}+1} \right] + (1-\beta) \,,
\end{equation}
where $\phi$ is the orbital phase defined from the superior conjunction of pulsar A, and the subscripts $i$ and $e$ refer to the ingress and egress, respectively, of the location ($\phi_{i,e}$) and steepness ($w_{i,e}$) parameters. Since the light curves were normalized such that the out-of-eclipse values are unity, we forced the out-of-eclipse level to be unity and kept the eclipse depth, $\beta$, as a free parameter. Hence, $\beta=0$ and 1 implies full transmission and absorption, respectively. In this framework, we can evaluate the full width at half-maximum as FWHM$=\phi_e - \phi_i$. Results are presented in Table \ref{t:fit_results}.

\begin{table}[ht!]
\caption{Double Fermi Function Fit Results\tablenotemark{a}}\label{t:fit_results}
\scriptsize
\begin{center}
\begin{tabular}{lcccccc}\hline
Date & $\phi_i$ & $\phi_e$ & $w_i$ & $w_e$ & $\beta$ & FWHM \\
{\scriptsize (MJD)} & {\scriptsize (deg)} & {\scriptsize (deg)} & {\scriptsize (deg)} & {\scriptsize (deg)} & {\scriptsize (Normalized)}\tablenotemark{b} & {\scriptsize (deg)} \\ \hline
\multicolumn{7}{c}{325\,MHz} \\ \hline
53191 & 0.52(4) & 0.90(0) & 0.22(3) & 0.01(0) & 0.96(3) & 1.42(4) \\
      & 0.61(5) & 0.91(1) & 0.29(4) & 0.05(1) & 1.00(4) & 1.53(5) \\
\hline
\multicolumn{7}{c}{427\,MHz} \\ \hline
53005 & 0.65(6) & 0.83(2) & 0.24(5) & 0.08(2) & 1.00(6) & 1.47(7) \\
      & 0.59(3) & 0.85(1) & 0.18(3) & 0.04(1) & 0.97(3) & 1.45(4) \\
      & 0.59(4) & 0.85(1) & 0.24(3) & 0.05(1) & 1.00(4) & 1.43(4) \\
\hline
\multicolumn{7}{c}{820\,MHz} \\ \hline
52992 & 0.55(5) & 0.75(2) & 0.23(3) & 0.07(1) & 0.99(5) & 1.30(5) \\
      & 0.58(3) & 0.75(1) & 0.21(3) & 0.05(1) & 1.00(3) & 1.34(4) \\
52997 & 0.62(3) & 0.76(1) & 0.20(3) & 0.06(1) & 0.96(3) & 1.38(4) \\
      & 0.64(3) & 0.76(1) & 0.22(3) & 0.06(1) & 0.99(3) & 1.40(4) \\
      & 0.62(3) & 0.73(0) & 0.14(2) & 0.01(0) & 0.91(3) & 1.36(3) \\
53211 & 0.70(3) & 0.79(2) & 0.10(3) & 0.04(1) & 0.92(3) & 1.49(4) \\
      & 0.57(6) & 0.78(2) & 0.19(5) & 0.06(2) & 0.98(6) & 1.35(6) \\
53311 & 0.61(5) & 0.79(1) & 0.14(4) & 0.01(1) & 0.88(5) & 1.39(5) \\
      & 0.56(6) & 0.76(2) & 0.21(5) & 0.06(2) & 1.06(7) & 1.32(6) \\
      & 0.54(16) & 0.73(3) & 0.37(10) & 0.07(3) & 1.13(16) & 1.27(16) \\
53379 & 0.69(5) & 0.79(3) & 0.16(4) & 0.10(3) & 0.96(5) & 1.48(6) \\
\tablenotemark{c} & \nodata & \nodata & \nodata & \nodata & \nodata & \nodata \\
      & 0.60(6) & 0.76(2) & 0.25(5) & 0.05(2) & 1.02(6) & 1.36(7) \\
\hline
\multicolumn{7}{c}{1400\,MHz} \\ \hline
52984 & 0.51(7) & 0.76(4) & 0.20(5) & 0.09(3) & 0.91(8) & 1.28(8) \\
      & 0.51(5) & 0.69(2) & 0.16(4) & 0.04(1) & 0.94(5) & 1.20(5) \\
\hline
\multicolumn{7}{c}{1950\,MHz} \\ \hline
53212 & 0.52(7) & 0.67(2) & 0.23(5) & 0.05(2) & 1.01(8) & 1.19(7) \\
      & 0.56(3) & 0.64(0) & 0.11(3) & 0.00(0) & 0.87(4) & 1.21(3) \\
53378 & 0.64(3) & 0.67(0) & 0.13(3) & 0.01(0) & 0.89(3) & 1.31(3) \\
      & 0.59(5) & 0.70(2) & 0.18(4) & 0.06(2) & 0.91(6) & 1.30(6) \\
\hline
\end{tabular}
\tablenotetext{1}{Numbers in parentheses are the $1\sigma$ uncertainties on the last significant digits.}
\tablenotetext{2}{Since the pulsed flux level is normalized to be unity in the out-of-eclipse region, depths $\beta$ of 0.0 and 1.0 indicate full transmission and total absorption, respectively.}
\tablenotetext{3}{Convergence problems with the non-linear least squares algorithm were encountered with the second eclipse of MJD 53379 and hence it was discarded from the eclipse duration and depth analysis.}
\end{center}
\end{table}

\begin{figure}
\centering
\includegraphics[width=0.95\columnwidth]{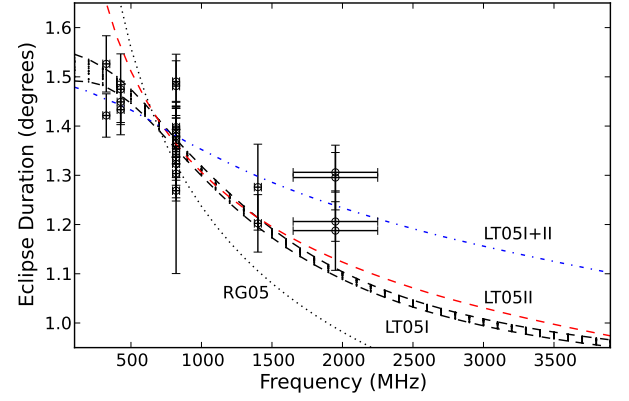}
\caption{Eclipse duration as a function of radio frequency for data collected between 2003 December and 2005 January. Here, the duration corresponds to the full width at half-maximum of a double Fermi function fitted to the eclipse profiles (Equation (\ref{eq:fermi})). The vertical error bars represent the $1\sigma$ uncertainties on the fits while the horizontal error bars indicate the receiver's bandwidth. The observed behavior of the duration is in good agreement with that calculated from light curves simulated using the \citet{lyu05a} synchrotron absorption/dipolar geometry model at the best-fit parameters from \citet{bre08a} --- LT05I (dashed black) use a constant electron density with a hard cutoff radius, while LT05II (dashed red) use an electron density that increases like a power law with radius. The spread in the theoretical predictions of LT05I is due to fact that the modulation features slightly change depending on the zero point of the rotational phase of pulsar B with respect to the orbital position. The dotted line shows the best-fit $\nu^{-1/3}$ relationship predicted by the RG05 model. The dash-dotted line shows a hybrid version of the LT05I and LT05II models, with a truncated power-law distribution (see Section~\ref{s:hybrid}).}
\label{f:duration}
\end{figure}

Figure \ref{f:duration} shows the eclipse duration as a function of radio frequency inferred from the FWHM for the data reported in Table \ref{t:fit_results}. Since the goodness-of-fit is difficult to assess given that we fitted a smooth function to profiles that have more complex structure, we verified that our inferred errors for the duration of the eclipses at 820\,MHz are consistent with the observed scatter of measured values at this frequency.\footnote{We chose 820\,MHz because it contains the largest number of data points.} We observe that the frequency dependence of the eclipse duration is clearly not steep enough to account for FWHM\,$\propto \nu^{-1/3}$ predicted by the RG05 model as introduced in Section~\ref{s:intro}. This model relies on the fact that the enhanced absorbing plasma density is provided by pulsar A's radio beam illumination and cyclotron absorption increases the trapping efficiency of electrons in the magnetosphere.

We compared the relationship between the eclipse duration and the radio frequency with the \citet{lyu05a} models which are based on synchrotron absorption from the relativistic electrons confined within the closed field lines of the magnetosphere. In the first version of the model, LT05I, a constant electron density distribution is assumed for the plasma inside the closed field lines, before dropping to zero for the field lines closing beyond the truncation radius $R_{\rm mag}$. \citet{bre08a} demonstrated the success of this model at reproducing the 820\,MHz eclipses. Because of the strong dependence of the eclipse profile on the geometry, we worked using the best-fit parameters found in \citet{bre08a} and hence the radio frequency was the only parameter to vary in the calculation of the eclipse duration. From Equation~(52) in \citet{lyu05a}, we can write the optical depth as
\begin{equation}\label{eq:lt05i}
	\tau_\nu = \mu \nu^{-5/3} \int_{-\infty}^{+\infty} I(x) (B \sin \kappa)^{2/3} \, dx \,,
\end{equation}
with
\begin{equation}
	I(x) = \left\{
		\begin{array}{ll}
		1 & \mbox{if $R_{\rm max}(x) < R_{\rm mag}$ and $\nu_{\rm B,e}/\nu > \langle \gamma_e \rangle \simeq 10$} \\
		0 & \mbox{otherwise}
		\end{array} \right. \,.
\end{equation}
Here $\kappa$ is the angle between the line-of-sight and the local magnetic field direction, $\mu$ is a constant that accounts for various physical quantities and plasma properties, $R_{\rm max}$ stands for the maximum extent of a field line, $R_{\rm mag}$ is the truncation radius of the magnetosphere, $\nu_{\rm B,e}$ is the local synchrotron frequency, and $\langle \gamma_e \rangle = 3 k_{\rm B} T_e/m_e c^2$.

As we can see in Figure \ref{f:duration}, the eclipse duration inferred from the LT05I model is able to reproduce the data fairly well, though the 1950\,MHz observations suggest a longer duration than expected. Given the poor S/N and the wide bandwidth in this band, it is not clear whether this difference is significant or not. In principle, data at higher radio frequencies ($\gtrsim 3000$\,MHz) would provide a good test but the fact that pulsar A becomes very faint beyond 2200\,MHz poses an observing challenge. The co-addition of several individual eclipses would improve the S/N but the process requires shifting the light curves so that the spin phases of pulsar B are aligned, as described in Section~\ref{s:periodicity}, in order to sum the modulation features coherently. We do not possess sufficient high-frequency data to attempt this and since pulsar B is now undetectable \citep{per10b}, it would be technically difficult, if not impossible, for eclipses that are not recorded in a continuous observation to be combined. A possible workaround, though not ideal, would be to conduct another multi-frequency campaign in the same bands as our current data but including much more 1950\,MHz observations in the hope of obtaining better statistics and perhaps recording some eclipses with low radio frequency interference.

We also examined the possibility of a different distribution for both the electron density and temperature. If one assumes that the electron density and temperature are constant along each field line, and are power-law functions of the distance (i.e., $T_e \propto R_{\rm max}^\alpha$ and $n_e \propto R_{\rm max}^\beta$), then using Equations (39) and (50) of \citet{lyu05a} one can write the optical depth as
\begin{equation}
	\tau_\nu = \mu^{\prime} \nu^{-5/3} \int_{-\infty}^{+\infty} R_{\rm max}(x)^{\left( \beta-5\alpha/3 \right)} (B \sin \kappa)^{2/3} \, dx \,,
\end{equation}
with $\mu^{\prime}$ being a constant that accounts for various physical quantities and plasma properties.

One readily recovers Equation~(\ref{eq:lt05i}) by assuming constant temperature and density distribution (i.e., fixing $\alpha = \beta = 0$, and imposing the truncation radius). On the other hand, the LT05II model (see Section~\ref{s:intro}) proposes that the electron density is given by \citep[][Equation~(56)]{lyu05a}
\begin{equation}\label{eq:density}
	n_e = \frac{k_B T_e}{m_p c^2} \frac{B^2}{2\pi m_p c^2} \,,
\end{equation}
which implies that $\beta = \alpha - 6$, with $\alpha$ suggested to be equal to 7 \citep[][Equation~(59)]{lyu05a}.

As we can see from Figure \ref{f:duration}, the behavior of the LT05II eclipse duration is for most parts very similar to that of the LT05I and is relatively consistent with the observations. This behavior was expected since the only explicit frequency dependence of LT05I and LT05II comes from the $\nu^{-5/3}$ component of the synchrotron absorption coefficient.

We note, however, the departure of the LT05II model at low frequency, where it produces eclipse durations that increase exponentially and are marginally longer than observed. This is probably a consequence of the very peaked electron power-law temperature distribution, which produces a runaway absorption of low-frequency radio photons at large magnetospheric radii. Finally, we also find that the eclipse duration is not very sensitive to a variation of the power-law index. This is mainly due to the constant $-6$ factor arising from the $B^2$ dependence in the electron density (see Equation~(\ref{eq:density})) of the LT05II model, which diminishes the importance of the power-law index $\alpha$.

\subsection{Depth}

\begin{figure}
\centering
\includegraphics[width=0.95\columnwidth]{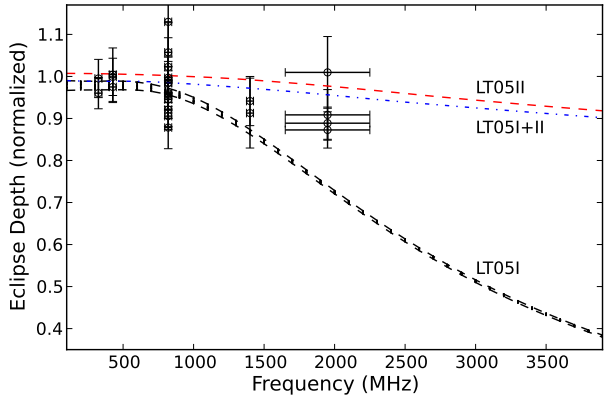}
\caption{Same as in Figure \ref{f:duration} but for the eclipse depth as a function of radio frequency for data collected between 2003 December and 2005 January. The depth corresponds to the parameter $\beta$ in Equation (\ref{eq:fermi}), which fits the eclipse light curve for a double Fermi function (see Table~\ref{t:fit_results}). The vertical error bars are the $1\sigma$ uncertainties on the fits and the horizontal error bars indicate the receiver's bandwidth. The dashed lines represent the prediction of two versions of the \citet{lyu05a} model, with a constant plasma density and sharp cutoff radius in black, LT05I, and with a radial power-law distribution of the electron density and temperature in red, LT05II. The dash-dotted line shows a hybrid version of the LT05I and LT05II models, with a truncated power-law distribution (see Section~\ref{s:hybrid}).}
\label{f:depth}
\end{figure}

Complementary to the eclipse duration in the investigation of the plasma properties is the radio frequency dependence of the eclipse depth, which we define as the $\beta$ parameter in Equation (\ref{eq:fermi}). Figure \ref{f:depth} suggests that there is little or no change in the eclipse depth as a function of frequency in the range $325-1950$\,MHz. This contrasts with the expected behavior of the hard cutoff model of LT05I which predicts a rough linear decrease of the depth between 1000 and 3500\,MHz. Conversely, the power-law distribution of the LT05II model displays much better agreement as its radio frequency evolution is very mild.

\section{A Hybrid Model}\label{s:hybrid}
The LT05I and LT05II models both display contradictory features. On the one hand, LT05I  agree fairly well with the duration of the eclipses, whereas LT05II predict durations that are too long in the low radio frequency regime. On the other hand, LT05II seem to account much better than LT05I for the very small variation of the eclipse depth as a function of frequency. The main conclusion that should be drawn is that the magnetosphere must possess a region where the optical depth is large over a wide range of radio frequencies before dropping abruptly in a very sharp transition region. It therefore appears that the actual distribution of temperature and density might be a compromise between the LT05I and LT05II models: a power-law which is truncated at a characteristic radius, i.e. $T_e \propto n_e \propto r^\alpha$ for $R_{\rm max}(r) < R_{\rm mag}$ and 0 otherwise. In this fashion, the optical depth would be sufficiently large to high-frequency radio photons traveling within the cutoff radius to produce deep eclipses even in this regime. At the same time, the truncation of the magnetosphere would limit the spatial extent of the eclipses at low frequencies where electrons have larger synchrotron absorption cross-sections to low-frequency photons over greater radii. Figures \ref{f:duration} and \ref{f:depth} show such a hybrid model. Without any attempt to optimize the power-law index and the truncation radius, it displays much better agreement to the data than the other models.

We note that this hybrid model shares similarities with Earth's magnetosphere which is immersed in the solar wind and possesses a sharp transition region, the plasmapause, where the plasma density drops by almost two orders of magnitude \citep{carp63a}. The sudden decrease of plasma density is due to the fact that confinement is efficient only where the field lines are near corotation. Instead of extending all the way to the limit of the light cylinder, as one would naively expect for an isolated radio pulsar, the plasmasphere of pulsar B is confined closer to the neutron star. As demonstrated by the particle-in-cell simulations of \citet{aro05a}, pulsar A's magnetized wind penetrates deep into the magnetosphere of pulsar B and reconnects with its magnetic field. This generates a tangential stress that breaks down corotation. We can compare the size of the plasmasphere that we obtain from our modeling with that estimated from the standoff distance determined by equating pulsar A's wind pressure to pulsar B's magnetic pressure. The inferred plasmasphere radius from our modeling \citep{bre08a,bret09a} is $R_{\rm mag} = 0.023\,D_{\rm AB}$, where $D_{\rm AB}$ is the orbital separation. This implies a physical size of $2 \times 10^9$\,cm, which is half that estimated using the standard magnetic dipole braking equation with fiducial neutron star parameters \citep{aro05a,lyu05a}. Such a conundrum is not impossible to resolve if pulsar B/A's magnetic field has a smaller/larger strength than estimated from the spin-down, and/or if further numerical simulations can show that the spin-down torque on pulsar B can be larger than that inferred for a dipole in vacuum. If arising from the torque only, this would require a factor $\sim 16$, which is not incompatible with the limit of $\sim 10$ estimated by \citet{lyu05a}.

From our modeling of the eclipses we observe that a rather large plasma density is required in pulsar B's magnetosphere, with an electron multiplicity factor $\sim 10^5$ the standard Goldreich$-$Julian density. One possibility to explain the high plasma density is that it is specific to the double pulsar system as a consequence of the interaction between pulsar B's magnetosphere and pulsar A's relativistic wind. ``Radial diffusion'' type effects such as those observed in planetary magnetospheres could bring a high particle influx inside the magnetosphere of pulsar B \citep{schu74a}. On the other hand, another possibility might be that a large electron multiplicity is instead a generic feature to all pulsars (e.g., using diffusion from the open field lines, for example; \citealt{aron79a}). More detailed theoretical would be required to study the viability of different mechanisms at reproducing large plasma density. However, without secure measurements for other sources it is difficult to draw conclusions for the pulsar population as a whole.

\section{Conclusion}\label{s:conclusion}
In this paper, we have presented the detection of flux modulations in the eclipse light curves of pulsar A at radio frequencies between 325 and 1950\,MHz, extending from the initial discovery of \citet{mcl04a} at 820\,MHz. This shows that this behavior is intrinsic to the eclipse mechanism. Furthermore, our quantitative analysis of the light curves reveals that the periodicity is indeed exactly related to the period and higher harmonics of pulsar B's rotation and only occurs within the orbital range of the eclipse. The flux modulations show a dynamic evolution, switching between modes dominated by power at the spin period of pulsar B and that of its first harmonic. Their phases and behavior are consistent with the absorbing material corotating with pulsar B and a dipolar magnetic field having a sharp optical depth transition region located in the closed field lines.

Regarding the properties of the absorbing plasma, we find that the RG05 model \citep{raf05a} produces an eclipse duration that is too radio-frequency sensitive. This disfavors the hypothesis that a only ``modest'' electron multiplicity ($\sim 100$) is reached from the illumination of pulsar A's radio flux and sustained in a steady state by the heating via resonant cyclotron absorption. Instead, our observations suggest that the plasma contained in pulsar B's magnetosphere possesses a ``large'' electron multiplicity ($\sim 10^5$).

Unfortunately, the density and temperature distributions of the relativistic electrons are more difficult to constrain. From our work, we conclude that for a wide range of radio photons the optical depth is very large up to a characteristic cutoff radius beyond which it drops abruptly. Such a behavior would be reproduced by a hybrid model of LT05I and LT05II for which the plasma electron temperature and density distributions are both truncated radial power-laws and resemble Earth's plasmasphere as suggested by \citet{aro05a}. The measured size of the plasmasphere $2 \times 10^9$\,cm is about half that of the expected magnetic field to wind pressure standoff distance and could imply that the spin-down of pulsar B is larger than expected from standard estimates. It is not clear whether the properties of pulsar B's magnetosphere can be generalized to other isolated pulsars since interactions with pulsar A's wind are certainly not negligible. Nevertheless, it represents one of the best available opportunity to study the immediate vicinity of a pulsar.

\acknowledgments

The National Radio Astronomy Observatory is a facility of the National Science Foundation operated under cooperative agreement by Associated Universities, Inc. R.P.B. thanks C. Thompson for insightful discussions about the eclipse mechanism. V.M.K. holds Lorne Trottier and Canada Research Chairs, and acknowledges support from NSERC, CIFAR, a Killam Research Fellowship and FQRNT. M.A.M. is supported by the WVEPSCOR, the Research Corporation, and NSF. Pulsar research at UBC is supported by an NSERC Discovery Grant. The authors also wish to thank the anonymous referee for pointing out the importance of the plasmasphere boundary in our paper.

{\it Facility:} \facility{GBT}.

\bibliography{apj0737_1}







\end{document}